\documentclass[aps, twocolumn, prl, showpacs,preprintnumbers,amsmath,amssymb,superscriptaddress]{revtex4-1}

\usepackage{amsmath,amsfonts,amssymb,graphics,graphicx,epsfig,color,times,indentfirst,layout,hyperref,subfigure,multirow,bm}
\usepackage{hyperref}
\usepackage{bbold}
\usepackage{bm}
\hypersetup{linktocpage,,colorlinks=true,citecolor=red}

\usepackage{soul}

\usepackage{graphicx}
\usepackage{dcolumn}
\usepackage{bm}
\usepackage{amssymb}
\usepackage{amsmath}
\usepackage{xcolor}
\usepackage{mathrsfs}
\usepackage{fixmath}
\hypersetup{colorlinks=true}
\usepackage[all]{hypcap} 

\newcommand{\mbd}{\mathbold}

\begin{document}

\title{Strange metal in paramagnetic heavy-fermion Kondo lattice: Dynamical large-$N$ fermionic multi-channel approach }
\author{Jiangfan Wang}
\affiliation{Department of Electrophysics, National Chiao-Tung University, Hsinchu 30010, Taiwan, R.O.C.}
\author{Yung-Yeh Chang}
\affiliation{Department of Electrophysics, National Chiao-Tung University, Hsinchu 30010, Taiwan, R.O.C.}
\author{Chung-Hou Chung}
\affiliation{Department of Electrophysics, National Chiao-Tung University, Hsinchu 30010, Taiwan, R.O.C.}
\affiliation{Physics Division, National Center for Theoretical Sciences, Hsinchu 30013, Taiwan, R.O.C.}
\date{\today}

\begin{abstract}
The mechanism of strange metal (SM) with unconventional charge transport near magnetic phase transitions has become an outstanding open problem in correlated electron systems. Recently, an exotic quantum critical SM phase was observed in paramagnetic frustrated heavy-fermion materials near Kondo breakdown. We establish a controlled theoretical framework to this issue via a dynamical large-$N$ fermionic multichannel approach to the two-dimensional Kondo-Heisenberg lattice model, where KB transition separates a heavy-Fermi liquid from fermionic spin-liquid state. With Kondo fluctuations being fully considered, we find a distinct SM behavior with quasi-linear-in-temperature scattering rate associated with KB. When particle-hole symmetry is present, signatures of a critical spin-liquid SM phase as $T\rightarrow 0$ are revealed with $\omega/T$ scaling extended to a wide range. We attribute these features to the interplay of critical bosonic charge (Kondo) fluctuations and gapless fermionic spinons. The implications of our results for the experiments are discussed.  
\end{abstract}

\maketitle

Over the recent decades, there has been growing experimental evidences whose thermodynamic and transport properties violate the Landau's Fermi liquid (FL) paradigm \cite{LFL-Landau}. These non-Fermi liquid (NFL) behaviors, ranging from unconventional superconductors \cite{Taillefer-AnnRev-HiTc, Petrovic-115} to Kondo quantum dots \cite{Gordon-2CK-Exp}, often exists close to a magnetic quantum phase transition (QPT) and exhibits ``strange metal (SM)" phenomena with (quasi-)linear-in-temperature resistivity and logarithmic-in-temperature singular specific heat coefficient \cite{Lohneysen-RMP}. More recently, an exotic NFL paramagnetic quantum critical SM state was experimentally found in geometrically frustrated heavy-electron Kondo lattice systems. Examples include CePd$_{1-x}$Ni$_x$Al ($0\leq x \leq 1$) on Kagome lattice \cite{zhang2018kondo,lucas2017entropy,TB-PD-CePdAl-PJSun, 2019-Sun-CePdAl, Lohneysen-CePdNiAl-PRB} and YbAgGe on triangular lattice \cite{schmiedeshoff2011multiple,tokiwa2013quantum,PRB-YbAgGe-NFL}
\begin{figure}[ht]
\centering
\includegraphics[width=0.38\textwidth]{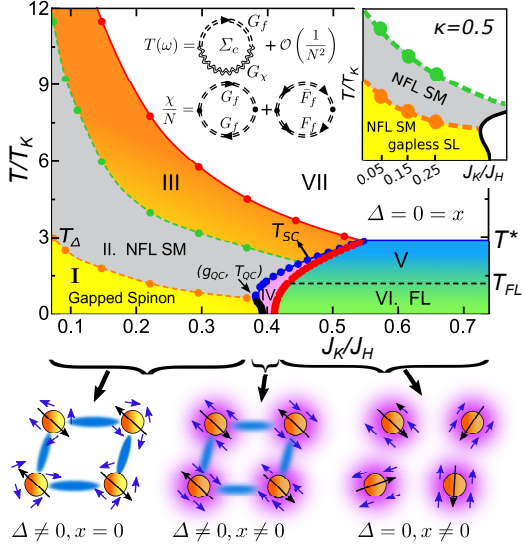}
\caption{Top: Finite-temperature phase diagram for $\kappa = 0.3$. 
Different colors correspond to regions with different behaviors of physical quantities. Regions I, II, and III are dominated by RVB, while regions V and VI refer to the Kondo-screened phase. Region IV is a coexisting (superconducting) phase, while region VII is the decoupled phase. The FL phase exists for $T< T_{FL}$ (marked by green), while the NFL SM region is bounded by the green and orange curves (marked by grey). $(g_{QC}, \, T_{QC})$ is the end-point of the crossover of region I.
Left insets show the  diagrams for the conduction electron $T$-matrix $T(\omega)$ and spin susceptibility $\chi(\omega)$ in the large-$N$ limit with solid, dashed and wavy lines refer to the conduction electron, pseudo-fermion and Kondo fluctuation fields, respectively. The right inset is the schematic phase diagram for $\kappa = 0.5$. The yellow region is a NFL gapless spin liquid (SL) phase, extended to $T\rightarrow 0$. The NFL SM region (gray area)  corresponds to region II of the main phase diagram. Bottom: Schematic plot of the phases on the Kondo lattice. Blue bonds and purple clouds refer to the RVB singlets and Kondo-screening clouds, respectively.}
\label{fig:PhaseDiag}
\end{figure}

In these materials, the geometrical frustration suppresses the long-range magnetic order of these antiferromagnetically (AF) coupled local spins and may lead to a spin-disordered paramagnetic spin-liquid state with fractional spin excitations (or spinons), while the observed large effective electron mass indicates the presence of Kondo screening between conduction and local electron spins. The competition between these two gives rise to rich phase diagram and novel QPT, which are largely unexplored. In particular, the microscopic mechanism of fermionic spin-liquid with SM feature still remains an outstanding issue \cite{zhang2018kondo}.

The well-known Doniach's framework-competition between the paramagnetic heavy Fermi liquid with Kondo correlation and antiferromagnetic long-range order-offers a qualitative understanding of many heavy-fermion systems close to antiferromagnetic Kondo breakdown (KB) quantum critical point (QCP), such as: YbRh$_2$Si$_2$ \cite{Paschen-Nat-2004, Gegenwart-Sci-2007, Friedemann-PNAS-2010}, 
CeCu$_{6-x}$Au$_x$ \cite{Schroder2000CeCuAu, Lohneysen-CeCuAu, Lohneysen1994CeCuAu, Lohneysen-XPS-CeCuAu, Lohneysen-Photoemission-CECUAU}. The extension of Dynamical Mean-Field Theory (DMFT) captures various aspects of this transition \cite{Si2001Nature-BFK-EDMFT, Si2003prb, EDMFT-Anisotropic-KL, QM-EDMFT-KL}. In the absence of long-ranged magnetic order, on the other hand, the controlled dynamical large-$N$ approaches of the Kondo-Heisenberg (KH) lattice model provide an appropriate theoretical starting point to capture spin liquid-to-Kondo QPT. Wherein the AF coupling is represented by the resonating-valence-bond (RVB) spin singlets and the heavy electrons via Kondo hybridization are described by the spin-charge separated spinons and holons \cite{coleman2005sum,Yashar1D, Komijani2018, Jiangfan-swb}. A dynamical Sp($N$) Schwinger boson approach \cite{Coleman-SWB-2006} was used to capture aspects of bosonic spin liquid near AF-KB QCP \cite{Komijani2018, Jiangfan-swb}, while a fermionic SU($N$) mean-field \cite{pixley2014quantum} and dynamical \cite{Burdin2002PRB} approach was applied to address the QPT of frustrated KH models.


 In this paper, we develop a distinct controlled approach: the multichannel pseudo-fermion dynamical large-$N$ [Sp($N$)$\times$ SU($K$) with $N$ and $K$ being the number of spin flavors and Kondo-screening channels] approach to the KH model on a square lattice, with Sp($N$) referring to the symplectic group symmetry.
The AF coupling is described by the fermionic spinons, while the Kondo hybridization is described by bosonic holon. 
This [Sp($N$)] approach is appropriate to describe the resonant-valence-bond (RVB) spin-liquid in magnetic systems on geometrically frustrated lattices, 
with the parameter $\kappa =K/N \leq 1/2$ measuring the effective spin moment per flavor or degree of quantum fluctuations \cite{Read-PRL-VBS,Read-PRL-FrustratedAF, Chung-PRB-SSL, Chung_JPCM_largeNHubbard}.
A similar approach in Ref. \cite{parcollet1998overscreened} was used to study the NFL state of a single impurity Kondo model with SU($N$)$\times$ SU($K$) symmetry, which does not allow for condensation of charged bosons to form Kondo singlets, and hence has difficulty describing the heavy-electron Kondo state.
We circumvent this difficulty by allowing for the channel symmetry to be spontaneously broken, leading to a condensation of the bosonic Kondo hybridization field and a fully screened-Kondo phase. The Kondo fluctuations and the spinons are then fully considered by solving a set of self-consistent Dyson-like equations.
This approach allows us to simultaneously explore both the Kondo-screened and fermionic spin-liquid phases as well as possible novel NFL properties due to their critical excitations, which goes beyond the existing approaches \cite{pixley2014quantum,Burdin2002PRB,coleman1989kondo, Senthil2003prl, Pepin-2005-PRL}.
Remarkably, we find a distinct NFL SM state near the KB QCP. When particle-hole symmetry is present, a critical spin-liquid SM phase is uncovered with $\omega/T$ scaling extended to a wide range. We attribute these features to the interplay of critical bosonic charge (Kondo) fluctuations and gapless fermionic spinons  
\cite{SWB-comparison}. 

\textit{Model- }We start from the Kondo-Heisenberg model on a 2D square lattice $H=\sum_{i}H_{0}(i)+\sum_{i}H_K(i)+\sum_{\left\langle ij\right\rangle}H_{J}(i,j).$ Here, the half-filled conduction electron reservoir is described by the sum of independent electron bath at site $i$ as: $H_0(i) = \sum_{\mbd{P},\alpha} \varepsilon_{\mbd{P}} c^\dagger_{i\alpha}(\mbd{P})c_{i\alpha}(\mbd{P})$ with $\alpha$ being the spin index, $\mbd{P}$ being the momentum, and $\varepsilon_\mbd{P}$ being the energy of the electron on the ``bath lattice" that is orthogonal to the impurity lattice. We assume each electron bath constitutes a single band with half band width $D$ and a constant density of states. 
Each local impurity spin $\mbd{S}_i$ can be screened by the conduction electron spin $\mbd{\sigma}_i = 1/2 \sum_{\alpha\beta} \psi^\dagger_{i\alpha} \mbd{\sigma}_{\alpha\beta} \psi_{i\beta}$ with $\psi_{i\alpha} =\sum_\mbd{P} c_{i\alpha}(\mbd{P})$ via $H_K(i) = J_K  \mbd{S}_i \cdot \mbd{\sigma}_i$. The local spins are coupled by the antiferromagnetic Heisenberg term $H_J(i,j) = J_H  \mbd{S}_i\cdot \mbd{S}_j$. The finite-temperature phase diagram of this model is determined by $J_K/J_H$ and the dimensionless temperature, $T/T_K$, where $T_K=De^{-2D/J_K}$ is the single impurity Kondo temperature.

\textit{Method- }We decompose the spin operator using the Abrikosov pseudo fermion representation, $\bm{S}_i=\frac{1}{2}\sum_{\alpha\beta}f_{i\alpha}^\dagger \bm{\sigma}_{\alpha\beta}f_{i\beta}$, subject to the constraint, $n_f(i)=\sum_{\alpha}f_{i\alpha}^\dagger f_{i\alpha}=2S$. Via Hubbard-Stratonovich transformation, $H_K$ and $H_{J}$ can be factorized as $H_K\rightarrow \frac{1}{\sqrt{2}}\sum_{\alpha}\chi_{i}f_{i\alpha}^\dagger\psi_{i\alpha}+\text{H.c.}$ $+\frac{|\chi_i|^2}{J_K}$, $H_{J}\rightarrow\frac{1}{\sqrt{2}}\sum_{\alpha}\Delta_{ij}\tilde{\alpha}f_{i\alpha}f_{j,-\alpha}+\text{H.c.}+\frac{|\Delta_{ij}|^2}{J_H}$,
where $\chi_i$ and $\Delta_{ij}$ are the Kondo hybridization field and the singlet RVB between adjacent spins, respectively. Here $\tilde{\alpha}=\text{sgn}(\alpha)$. We further generalize the model to its multichannel large-$N$ limit, by replacing $\psi_{i\alpha}$ ($\chi_{i}$) with $\psi_{ia\alpha}$ ($\chi_{ia}$), where $a=1,2,\cdots,K$ denotes channel and $\alpha=\pm 1,\cdots, \pm N/2$ is the spin index. The large-$N$ limit is taken by $N,\,K \rightarrow \infty$ while the ratio $K/N = 2S /N =\kappa$ is kept fixed. This generalized large-$N$ model shows a total symmetry of Sp($N$)$\times$SU($K$). Note that $K=2S$ ensures the complete Kondo screening \cite{Coleman-SWB-2006}.

To describe the Kondo-screened heavy-electron Fermi-liquid state, we allow the Kondo hybridization field $\chi_{ia}$ to get Bose-condensed and break the SU($K$) symmetry: $\chi_{ia}(\tau)\rightarrow\sqrt{N}x\delta_{a,K}+\chi_{ia}(\tau)\left(1-\delta_{a,K}\right).$ The uniform and static Bose-condensed mean-field amplitude $x$, describing the Kondo singlet between the impurity and the conduction electron spins for the $K$-th channel, is determined by the saddle-point equations.
The remaining $K-1$ components of the $\chi$ fields are considered as fluctuating variables, which plays an important role in the development of NFL state.
Integrating out the $K$-th channel of the electrons, the effective action $\mathcal{S}$ of the remaining $K-1$ channels is given by
\begin{eqnarray}
\mathcal{S}&=&-\sum_{a\alpha}\psi_{a\alpha}^{*}G_{c0}^{-1}\psi_{ a\alpha}-\sum_{\alpha}f_{\alpha}^{*}\left(i\omega+\lambda-|x|^{2}G_{c0}\right)f_{\alpha}  \notag \\
 & &+\left[\frac{1}{\sqrt{N}}\sum_{a\alpha}\chi_{a}f_{\alpha}^{*}\psi_{a\alpha}+\Delta^*\sum_{\alpha}\tilde{\alpha}f_{\alpha}f_{-\alpha}\xi_{p}+c.c.\right] \notag \\
& & +\sum_{a}\frac{\left|\chi_{a}\right|^{2}}{J_{K}}+\beta NN_{s}\left(\frac{2\left|\Delta\right|^{2}}{J_{H}}+\lambda\frac{K}{N}+\frac{|x|^{2}}{J_{K}}\right) 
\label{Action}
\label{S_eff}
\end{eqnarray}
In Eq. (\ref{Action}), $\lambda$ is the uniform Lagrange multiplier that imposes the local constraint $n_f(i)=K$, $\Delta=\Delta_{i,i+\hat{x}}=\Delta_{i,i+\hat{y}}$ is the isotropic RVB, $G_{c0}(i\omega)=\sum_{\mbd{P}}(i\omega-\varepsilon_{\mbd{\mbd{P}}})^{-1}$ is the bare Green's function of electron baths diagonal in the spin subspace, $\xi_{\mbd{p}}=\cos p_{x}+\cos p_{y}$ is the dispersion of spinons. In the $\text{SU}(2)\cong \text{Sp}(2)$ limit ($\kappa =1/2$),  $\chi_i = \sqrt{2} x_i+ \tilde{\chi}_i$ with $\tilde{\chi}_i$ being the fluctuating Kondo field, keeping Eq. (\ref{Action}) the same form. With fluctuations of the Kondo hybridization included, our results go beyond the mean-field approaches \cite{pixley2014quantum,Burdin2002PRB,coleman1989kondo, Senthil2003prl}.


In the large-$N$ limit, Eq. (\ref{Action}) is solved by a set of self-consistent equations, similar to that within the non-crossing approximation (NCA) \cite{hewson1997kondo}. Due to the independent electron bath, the $\chi$ field is completely local. As a result, the local Kondo interaction leads to momentum-independent self-energies, and only the momentum-integrated Green's functions are involved \cite{Supple}: $G_{\chi}^{-1}(i\nu)=\left[-J_{K}^{-1}-\Sigma_{\chi}(i\nu)\right]$ and $G_f (i\omega) = \sum_\mbd{p}G_f (\mbd{p},i\omega)$, where $G_f (\mbd{p},i\omega) = \gamma(-i\omega)/[\gamma(i\omega)\gamma(-i\omega)+4\Delta_\mbd{p}^2]$ with
 $\gamma(i\omega)\equiv i\omega+\lambda-|x|^{2}G_{c0}(i\omega)-\Sigma_{f}(i\omega)$ and $\Delta_\mbd{p} \equiv \Delta \xi_\mbd{p}$ being the form factor of the extended $s$-wave pairing.
 Note that this local approximation is capable of effectively capturing important aspects of the local KB QCP \cite{Si2001Nature-BFK-EDMFT, Komijani2018,Jiangfan-swb}. By neglecting the $\mathcal{O}(1/N)$ vertex corrections \cite{Supple}, the leading $\mathcal{O}(1)$ Dyson-Schwinger equations for the self-energies read
$\Sigma_{\chi}(i\nu)=\sum_{\omega}G_{f}(i\omega+i\nu)G_{c0}(i\omega), \,\, \Sigma_{f}(i\omega) =-\kappa\sum_{\nu}G_{\chi}(i\nu)G_{c0}(i\omega-i\nu).
\label{eq:self-energy}
$
The self-energy of the $c$-electrons is of order $\mathcal{O}(1/N)$, and is therefore neglected in our calculations. Minimizing the free energy with respect to $\lambda$ enforces the local constraint of the local $f$-electron, to $\Delta$ gives the relation $\Delta(J_H,\,J_K,\,x,\,\Delta)$, and to $x$ determines $x(J_H,\,J_K,\,x,\,\Delta)$. We solve the the Green's functions and self-energies subject to constraints self-consistently.

\begin{figure}
\begin{centering}
\includegraphics[width=0.42\textwidth]{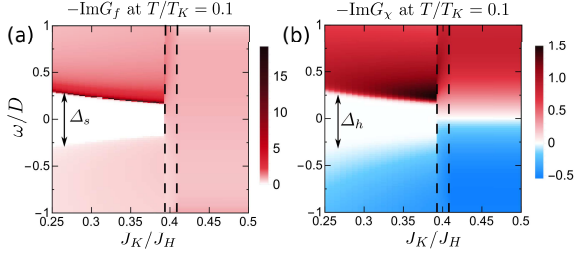}\par
\end{centering}
\caption{(a),(b) Density plot of the imaginary part of $G_f(\omega)$ and $G_\chi(\omega)$ at temperature $T/T_K=0.1$ for $\kappa = 0.3$. Inside the narrow region between the dashed lines is the coexisting phase where both $x$ and $\Delta$ are non zero. Both the $f$ and $\chi$ field open a gap at $J_K/J_H<0.39$. $\Delta_s$ and $\Delta_h$ indicate the sizes of the spinon and holon gaps, respectively.
 }
 \label{fig:GF}
\end{figure}

\textit{Finite-temperature Phase diagram- }
The finite-temperature phase diagram is shown in Fig. \ref{fig:PhaseDiag} with the choice of $\kappa = 0.3$ and $J_K/D=1.0$ ($T_K/D = 0.135$). The RVB spin-liquid metal phase exits for small values of $J_K/J_H$ ($\Delta\neq 0$, $x=0$, regions I,II, III), while the Kondo screened paramagnetic heavy-electron phase prevails at large $J_K/J_H$ ($\Delta=0$, $x\neq 0$, regions V, VI). A coexisting phase, an extended $s$-wave superconducting phase when electron baths are connected, is found at intermediate $J_K/J_H$ ($\Delta\neq 0$, $x\neq 0$, region IV) \cite{Chang-SSc-PRB}. A high temperature decoupled phase is reached when $\Delta=x=0$ (region I). 



The $T^\ast$ (blue) line sets the boundary between $x=0$ and $x\neq 0$. In the FL phase, $T^*$ corresponds to the mean-field Kondo coherence temperature below which the Bose-condensed Kondo hybridization develops phase coherence over the lattice,
with a value $T^*\sim2.8\,T_K$, consistent with experimental observation that $T_K \ll T^\ast$ \cite{Maple-arxiv-2017-Tstar}. 
At lower temperatures $T<T_{FL}\approx 1.2\, T_K$, the system becomes a Fermi liquid where specific heat coefficient reaches a constant at $T \sim T_{FL}$ \cite{Supple}.
At $J_K/J_H < g_c \approx 0.39$ and at low temperature (below the orange dashed line of Fig. \ref{fig:PhaseDiag}), the system develops gaps ($\Delta_s$, $\Delta_h$) in both the spinon and holon spectral functions where the thermodynamical observables and transport show an exponential decay as $T\rightarrow 0$ (see below) \cite{Supple}. Remarkably, for temperatures above the spinon gap, a NFL SM region (the gray area in Fig. \ref{fig:PhaseDiag}) is found, characterized by a linear-in-temperature dependence of the scattering $T$-matrix (see below).

\begin{figure}[t]
\begin{centering}
\includegraphics[width=0.45\textwidth]{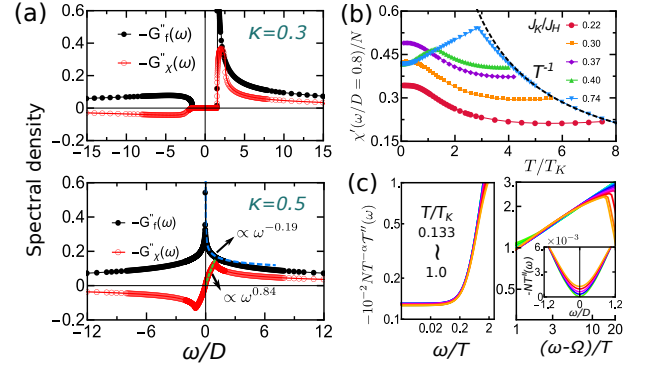}
\par
\end{centering}
\caption{(a) Spectral density: $-G_f^{\prime\prime}(\omega)$ and $-G^{\prime\prime}_\chi(\omega)$ for $\kappa=0.3$ (top) and $\kappa = 0.5$ (bottom) with $J_K/J_H=0.052$ and $T/T_K = 0.05$ fixed. (b) The real part of the dynamical spin susceptibility $\chi^\prime (\omega, \, T)$ versus temperature of local moments with $\kappa = 0.3$ and fixed frequency $\omega/D=0.8$ for different values of $J_K/J_H$. At high temperatures, $\chi^\prime (\omega, \,T)$ exhibits a Curie-law dependence $\chi^\prime \sim 1/T$ (black dashed line). (c) $\omega/T$ scaling of $T$-matrix with $T/T_K \in [0.133, \, 1.0]$ for $\kappa=1/2$ for (right) $\omega/T<1$ with $\alpha \sim 1.7$, and for (left) $\omega/T >1$ with $\alpha = 0.37$. $\Omega$ is a fitting parameter. Inset shows the unscaled $-N\mathcal{T}^{\prime\prime} (\omega)$.}
\label{fig:FL_NFL} 
\end{figure}

\textit{Green's functions- }Figures \ref{fig:GF} and \ref{fig:FL_NFL}(a) shows spectral weight of the $f$ and the $\chi$ fields at $\kappa = 0.3$. The density plot of the spectral functions at temperature $T/T_K=0.1$ shows a gap at $J_K/J_H<g_c$. The gap closes abruptly at $J_K/J_H = g_c$, indicating a first order KB transition. This first-order transition is identical to the metal-to-superconductor transition described in Ref. \cite{Ye-CuO}. The NFL SM state is linked to the end-point $(g_{QC},\,T_{QC})$ of this finite-temperature first-order transition line [the black dotted curve in Fig. \ref{fig:PhaseDiag}]. This suggests a quantum-critical end point associated with $g_{QC}$ when $T_{QC}$ is suppressed to zero \cite{2002-PRL-Millis-MetaQCP, schmiedeshoff2011multiple}.
At $\kappa=1/2$, however, the spinon gap in the RVB state vanishes [Fig. \ref{fig:FL_NFL}(a)], indicating an exotic gapless fermionic spin-liquid state (see below).

\textit{Dynamical spin susceptibility-} Fig. \ref{fig:FL_NFL}(b) shows the temperature dependence of the real part of the uniform dynamical spin susceptibility $\chi(\omega)$ of the local moments (see inset of Fig. \ref{fig:PhaseDiag} and Ref. \cite{Supple}).
On the FL side and at high temperatures ($\Delta=0=\chi$), the dynamical spin susceptibility $\chi^\prime(\omega) \equiv \text{Re}[\chi(\omega)]$ at a fixed frequency shows a $1/T$ Curie law of local spin moment; while it develops a kink at $T=T^*$ and decreases for $T< T^*$ to a saturated value as $T \rightarrow 0$ due to the formation of Kondo singlet.  
Interestingly, $\chi^\prime(\omega > \Delta_s)$ shows an increase with decreasing temperatures and tend to saturate at low temperatures, a signature of fermionic  spin-liquid \cite{Balents-2010-Nature-SL}. 
\begin{figure}[t]
\centering
\includegraphics[width=0.45\textwidth]{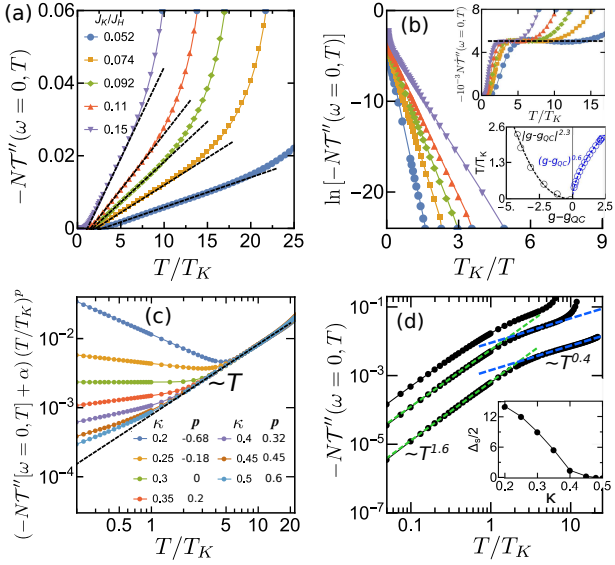}
\caption{(a) displays  $-N\mathcal{T}^{\prime\prime}(\omega = 0,\,T)$ with different values of $J_K/J_H$ at $\kappa = 0.3$. (b) shows the exponential decay of static $T$-matrix at low temperatures. Inset of (b): (top) temperature derivative of $T$-matrix, (below) power-law vanish of crossover scales with respect to $T_{QC}$: $T_\Delta - T_{QC} \propto |g-g_{QC}|^{r} (r \sim 2.3)$ (black dashed line), $T_{SC}-T_{QC} \propto |g-g_{QC}|^{s} (s\approx 0.6)$ (blue dashed line), associated with the gapped spinon and coexisting superconducting phases, respectively. (c) The scaling of static $T$-matrix for $J_K/J_H = 0.052$, where $\alpha$ and $p$ are non-universal $\kappa$-dependent constants. (d): $T$-matrix at $\kappa = 1/2$ for $J_K/J_H = 0.26$ (top), $0.15$ (middle), and $0.052$ (below) shows the same $T$-sub-linear and $T$-super-linear power laws in corresponding two temperature regimes (blue and green dashed lines). Inset of (d): the evolution of the gapped-spinon crossover with different values of $\kappa$ for fixed $J_K/J_H = 0.052$. The black dashed lines in (a) and (c) are linear fits.}
\label{fig:Chi}
\end{figure}

\textit{$T$-matrix and the strange metal state- }
The $T$-matrix of conduction electron is defined as $\mathcal{T}(\omega) = \Sigma_c(\omega)/[1-G_{c0}(\omega)\Sigma_c(\omega)]$, which reduces to $ \Sigma_c(\omega) \sim \mathcal{O}(1/N)$ in the $N \rightarrow \infty$ limit [inset of Fig. \ref{fig:PhaseDiag}].
Figure \ref{fig:Chi}(a) shows the imaginary part of static $T$-matrix, corresponding to the scattering rate, $\tau^{-1}(\omega=0,T) = -N\mathcal{T}^{\prime\prime}(\omega = 0, \, T)$, versus temperature for $\kappa = 0.3$. It displays a linear-in-$T$ NFL SM signature in the intermediate temperature range (region II of Fig. \ref{fig:PhaseDiag}), see the fitted dashed-lines. This NFL region shows quantum critical behavior associated with the KB transition at $(g_{QC},\,T_{QC})$ via an universal scaling form of $-N\mathcal{T}''(\omega=0,T)$, indicated by the scaling of the constant (flat) temperature derivative $-d/dT [N\mathcal{T}^{\prime\prime}(\omega=0,\,T)]$ and the power-law-in-$|g-g_{QC}|$ vanishing crossover scales (with respect to $T_{QC}$) on both sides of the transition, below which the NFL state disappears [insets of Fig. \ref{fig:Chi}(b)].
  In the gapped spin-liquid region, however, the $T$-matrix decays exponentially [Fig. \ref{fig:Chi}(b)].
  More generally, as $\kappa$ increases, the holon and spinon gaps are suppressed [inset of Fig. \ref{fig:Chi}(d)]. Strikingly, the static $T$-matrix in this region clearly shows a novel quasi-linear-in-$T$ power-law scaling [Fig. \ref{fig:Chi}(c)], i.e. $-N\mathcal{T}^{\prime\prime}(\omega=0,T)\sim T^{1-p(\kappa)}$ with $0<|p(\kappa)|<1$ over a wide range in low temperatures. The change from $T$-super-linear to $T$-sub-linear power-law in $T$-matrix with increasing $\kappa$ comes as a result of enhanced electron scattering at low temperatures via Kondo fluctuations with softened gaps. Similar quasi-linear power-law feature is also found in the dynamical $T$-matrix: $-N\mathcal{T}''(\omega,T=0)\propto (\omega - \Delta_s)^{1-\bar{p}(\kappa)}$ with $\bar{p} \approx p$  \cite{Supple,Chang-SM-PRB}. We attribute this SM feature to the critical Kondo fluctuations coupled to gapless fermionic spinon, manifested by the quasi-linear-in-$\omega$ power-law behavior in holon and spinon spectral functions outside the gaps [see Fig. \ref{fig:FL_NFL}(a)].

Surprisingly, at $\kappa=1/2$ where the spinon is half-filled, the spinon and holon gaps vanish due to particle-hole symmetry of the Hamiltonian [Fig. \ref{fig:FL_NFL}(a)]\cite{Supple}. As a result, we find a power-law singular (pseudogap vanishing) spinon (holon) spectral function $G_f\propto \omega^{-\bar{\alpha}} \, (G_\chi \propto \omega^{\bar{\beta}})$ as $\omega \rightarrow 0$, respectively [Fig. \ref{fig:FL_NFL}(a) and inset of Fig. \ref{fig:Chi}(d)], indicating an exotic Kondo-assisted gapless (critical) fermionic spin liquid phase \cite{2019-Sun-CePdAl, Algebraic-SL-2}. Remarkably, via the interplay of the gapless power-law spinon and holon spectral functions, the static $T$-matrix shows
a distinct SM feature with super-linear-in-$T$ power-law behavior in  $-N\mathcal{T}''(\omega = 0,T)\propto T^{1+p}$ as $T\rightarrow 0$ ($p\approx 0.6$) [Fig. \ref{fig:Chi}(d)] over a wide range in $J_K/J_H$ and $T/T_K$ \cite{Senthil-2004-PRB-NFL}, signature of a NFL SM phase (see phase diagram in inset of Fig. \ref{fig:PhaseDiag})  \footnote{Similar $T$-super-linear power-law in resistivity ($\rho \sim T^{4/3}$) was found in the SU($N$) approach to 2D KH model but is hindered by the leading $(\ln T)^{-1}$ suppression contributed from the U($1$) gauge fluctuations \cite{Senthil-2004-PRB-NFL}.}.
This power-law
exponent is well accounted for by power counting of spinon and holon spectral functions in $-N\mathcal{T}^{\prime\prime}(\omega=0, T)\approx \int (d\nu/\pi) \text{csch} (\beta \nu)G^{\prime\prime}_f(\nu)G_\chi^{\prime\prime}(\nu) \sim T^{1+\bar{\alpha}-\bar{\beta}}$ with $p \approx \bar{\alpha}-\bar{\beta}$, see Ref. \cite{Supple}. The striking similar NFL SM feature seen in the static and dynamical $T$-matrix is further supported by its $\omega/T$ scaling: $-T^{-\alpha} N \mathcal{T}^{\prime\prime}(\omega,T)\propto (\omega/T)^\alpha$ for $\omega<T$ and $\omega>T$, respectively  [see Fig. \ref{fig:FL_NFL}(c)]. By contrasting the critical (algebraic) spin-liquid with power-law correlations realized in frustrated magnets \cite{Algebraic-SL-2}, here we find a new example of such kind with SM feature stablized by the Kondo correlation. 

When the local electron baths are connected, this SM feature in $T$-matrix indicates a quasi-linear-in-$T$ resistivity $\rho $ via conductivity $\sigma (T) =1/\rho (T) \sim \int \tau(\omega)\partial_\omega  f(\omega)  d\omega$ with $f(\omega)$ being the Fermi function \cite{Chang-SM-PRB,Chang-SSc-PRB}; and we expect the Fermi liquid limit,  $-N\mathcal{T}''(\omega=0,T) \propto T^2$ as $T\rightarrow 0$, on the FL phase is recovered. Since the $T$-matrix is proportional to the local density of states, it can be directly compared to the low-temperature STM measurement.

\textit{Discussions-} Our results are relevant for describing CePd$_{1-x}$Ni$_x$Al with a reduced spin moment $s \approx 1/3$ per Ce site due to geometrical frustration. Under field and pressure, the system undergoes a KB transition from a paramagnetic SM state to a heavy Fermi liquid state. 
The $J_K/J_H$ ratio is expected to increase with increasing field or pressure \cite{Chang-SM-PRB,Chang-SSc-PRB}. 
The calculated SM feature in $T$-matrix ($\kappa = 1/2$) and the fermionic spin-liquid dynamical spin susceptibility ($\kappa < 1/2$) are qualitatively consistent with the quasi-linear-in-$T$ resistivity persistent to the lowest temperature, observed in its Ni-doped form \cite{lucas2017entropy}, as well as the susceptibility measurement for its pure form \cite{TB-PD-CePdAl-PJSun,zhang2018kondo,2019-Sun-CePdAl}, respectively.
 We have checked that our results are robust against finite-$N$ and finite-$K$ fluctuations with the inclusion of $\mathcal{O}(1/N)$ corrections \cite{Jiangfan-swb,Supple,foot-SLstability}. By introducing a weak hoping term between conduction baths, our RVB spin-liquid state is stable against singular U($1$) gauge fluctuations \cite{Senthil-2004-PRB-NFL} as the local U($1$) gauge symmetry is broken down to $Z_2$ by direct hoping of femionic spinons generated by the Kondo fluctuations \cite{Supple}. Consequently, our results can be readily generalized to frustrated lattices with the same $Z_2$ gauge symmetry. Our results open up an exciting new possibility of realizing quantum critical non-Fermi liquid phase in correlated electron systems. 

\textit{Acknowledgement- }The authors acknowledge discussions with A. M. Tsvelik, Q. M. Si, and S. Kirchner. This work is supported by the MOST (Grant NO.: 104-2112-M-009-004- MY3 and 107-2112-M-009-010-MY3), the NCTS of Taiwan, R.O.C. (C.-H. C.).
 
\bibliographystyle{apsrev4-1}
\bibliography{myref.bib}

\begin{thebibliography}{53}
\bibitem [{foo()}]{foot-SLstability}%
  \BibitemOpen
  \href@noop {} {}\bibinfo {note} {The gapless spin-liquid phase for
  $\kappa=1/2$ at large-$N$ is likely unstable against AF long-range order in
  the SU($2$) limit. Nevertheless, we expect this NFL SM behavior to survive at
  finite temperatures here and on certain frustrated lattices where critical
  spin-liquid is stable \cite{Algebraic-SL-2}.}\BibitemShut {Stop}%
\end{thebibliography}

\end{document}